\documentclass[superscriptaddress,showpacs,twocolumn,pre,floatfix]{revtex4}
\usepackage[dvips]{graphicx}
\usepackage[dvips]{color}
\usepackage{amsmath}
\usepackage{amssymb}

\newcommand{\kap}{\boldsymbol{\kappa}}
\newcommand{\rb}{{\bf r}}
\newcommand{\vb}{{\bf v}}
\newcommand{\bp}{\boldsymbol{\partial}}
\newcommand{\vct}[1]{\mathbf{#1}}

\newcommand{\D}{\boldsymbol{D}}

\begin{document}
\title{
Density profiles of a colloidal liquid at a wall under shear flow
}

\date{\today}
\author{J.M. Brader}
\affiliation{Department of Physics, University of Fribourg, CH-1700 Fribourg, Switzerland}
\email{joseph.brader@unifr.ch}
\author{Matthias Kr\"uger}
\affiliation{Fachbereich Physik, Universit\"at Konstanz, 78457 Konstanz, Germany}
\altaffiliation[Present address:]{ Department of Physics, Massachusetts Institute of Technology, Cambridge, Massachusetts 02139, USA}
\email{kruegerm@mit.edu}
\begin{abstract}
Using a dynamical density functional theory we analyze the density profile of a colloidal liquid near a wall under shear flow. Due to the symmetries of the system considered, the naive application of dynamical density functional theory does not lead to a shear induced 
modification of the equilibrium density profile, which would be expected on physical grounds.  
By introducing a physically motivated dynamic mean field correction we incorporate the missing shear induced interparticle forces into the theory. 
We find that the shear flow tends to enhance the oscillations in the density profile of hard-spheres 
at a hard-wall and, at sufficiently high shear rates, induces a nonequilibrium transition to a steady state characterized by planes of particles parallel to the wall. 
Under gravity, we find that the center-of-mass of the density distribution increases with shear rate, i.e., shear increases the potential energy of the particles.
\end{abstract}

\pacs{82.70.Dd, 83.80.Hj, 05.70.Ln, 71.15.Mb 
}
\keywords{density functional, diffusion, nonequilibrium}

\maketitle
\section{Introduction}
Classical Density Functional Theory (DFT) provides a powerful and general framework for determining 
the equilibrium microstructure and thermodynamics of classical many particle systems \cite{bob_advances,bob_review}. 
Of particular interest is the one-body density profile $\rho(\rb)$ resulting from application of a 
time-independent external potential field $V^{\rm ext}(\rb)$. 
Within DFT, the density profile of a one-component system follows from functional minimization of the Grand potential 
\begin{eqnarray}
\Omega[\,\rho\,]=\mathcal{F}_{\rm id}[\,\rho\,] + \mathcal{F}_{\rm ex}[\,\rho\,] + 
\int \!d\rb\, (V^{\rm ext}(\rb) - \mu)\rho(\rb), 
\end{eqnarray}
with respect to $\rho(\rb)$, where $\mu$ is the chemical potential and $\mathcal{F}_{\rm ex}[\,\rho\,]$ is the unknown `excess' part of the Helmholtz free energy, containing details of the interparticle interactions.  
The ideal gas free energy is given exactly by
\begin{eqnarray}
\mathcal{F}_{\rm id}[\,\rho\,]=\int\!d\rb \rho(\rb)[\ln(\Lambda^3\rho(\rb))-1], 
\end{eqnarray}
where $\Lambda$ is the thermal de Broglie wavelength. 
For many important model systems (e.g. hard-spheres \cite{rosenfeld}, colloid-polymer mixtures 
\cite{colpol,colpol1}, rod-sphere mixtures \cite{onsager}) 
there exist accurate approximations for the excess Helmholtz free energy which yield equilibrium 
density profiles in excellent agreement with those obtained from numerical simulation.

Given that DFT provides a clear method for determining equilibrium density distributions, it is natural to 
next investigate the {\em dynamics} of the density profile in the presence of a time-dependent external field 
$V^{\rm ext}(\rb,t)$. 
The simplest, phenomenological, route to an equation of motion for $\rho(\rb,t)$ is to assume that the average particle current 
${\bf j}(\rb,t)$ arises from the gradient of a nonequilibrium chemical potential 
\begin{eqnarray}
{\bf j}(\rb,t) = -\Gamma\rho(\rb,t)\nabla\mu(\rb,t), 
\end{eqnarray}
where $\Gamma$ is the mobility. 
Assuming that $\mu(\rb,t)$ is given by the functional derivative of the Helmholtz free energy with respect to 
$\rho(\rb,t)$ and employing the continuity equation thus leads to the familiar equation of dynamical density functional theory (DDFT) 
\begin{eqnarray}
\frac{\partial \rho(\rb,t)}{\partial t} = \nabla\cdot\left[ \Gamma\rho(\rb,t)
\nabla\frac{\delta \mathcal{F}[\rho(\rb,t)]}{\delta \rho(\rb,t)} \right],
\label{ddft}
\end{eqnarray}
where $\mathcal{F}$ is the equilibrium Helmholtz free energy functional, evaluated using the instantaneous 
nonequilibrium density profile. 
Although equation (\ref{ddft}) was first proposed by Evans \cite{bob_advances}, subsequent work has clarified greatly the nature of the approximations involved. Both Marconi and Tarazona \cite{marconi1,marconi2}, and 
Archer and Evans \cite{archer}, have demonstrated that approximating the nonequilibrium chemical potential using 
the equilibrium Helmholtz free energy is equivalent to assuming that the inhomogeneous pair correlations in nonequilibrium are the same as those for an equilibrium fluid with density profile $\rho(\rb,t)$. 
Specifically, for a system interacting via a pair-potential $\phi(|\rb_1-\rb_2|)$ the equilibrium sum-rule 
\cite{bob_advances}
\begin{eqnarray}\label{sumrule}
\int\!d\rb_2\, \rho^{(2)}(\rb_1,\rb_2)\nabla_1\phi(|\rb_1\!-\!\rb_2|) 
= \rho(\rb_1)\nabla_1\frac{\delta \mathcal{F}_{\rm ex}}{\delta \rho(\rb_1)},
\end{eqnarray} 
is assumed to hold. 
The integral on the l.h.s. of (\ref{sumrule}) occurs when coarse graining the $N$-particle Smoluchowski 
equation to the one-body level by integration over $N\!-\!1$ degrees of freedom. 
Applying the equilibrium equality (\ref{sumrule}) to close the resulting nonequilibrium expression leads 
directly to (\ref{ddft}). The implicit `adiabatic' approximation in applying (\ref{sumrule}) to 
nonequilibrium is that the one-time pair correlations are instantaneously equilibrated to those of an equilibrium system with density $\rho(\rb,t)$. 
For a wide range of systems, the good qualitative agreement between the results of Brownian dynamics simulation 
and DDFT validates the adiabatic approximation when applied to inhomogeneous fluid states out of equilibrium. 
However, the approximation may break down for dense fluids close to dynamical arrest (e.g. hard-sphere-like colloidal glasses), for which the structural relaxation time becomes large.

The possibility of going beyond the adiabatic approximation has been explored on a formal level \cite{finken}.   
However, an explicit and implementable method of incorporating temporal nonlocality into the 
theory remains to be found. 
More recently, the DDFT (\ref{ddft}) has been rederived using projection operator methods \cite{espanol} and 
extended to incorporate pair hydrodynamics \cite{rex1,rex2,rauscher_hydro}, orientational degrees of freedom \cite{rex3} 
and even self-propelled particles \cite{wensink}. Going beyond overdamped Smoluchowski dynamics, Marconi and 
coworkers have developed a DDFT-type approach to treat inertial systems which makes possible the study of 
thermophoresis \cite{marconi3}. 

In order to address the influence of external flow on inhomogeneous density 
distributions, a DDFT incorporating the advection of particles by a flowing solvent has been developed 
\cite{rauscher2}. 
The resulting advected-DDFT equation of motion has a form identical to that of the standard theory 
(\ref{ddft}), but with the time derivative replaced by the Stokes derivative 
\begin{eqnarray}
\frac{\partial \rho(\rb,t)}{\partial t} &+& \nabla\cdot(\rho(\rb,t)\vct{v}(\rb,t)) = \notag\\
&&\hspace*{0.5cm}\nabla\cdot\left[ \Gamma\rho(\rb,t)
\nabla\frac{\delta \mathcal{F}[\rho(\rb,t)]}{\delta \rho(\rb,t)} \right],
\label{adv_ddft}
\end{eqnarray}
where $\vct{v}(\vct{r},t)$ is the velocity of the solvent.
The advected-DDFT is therefore simply standard DDFT in the comoving frame and is thus subject to the same 
adiabatic approximation as the original theory. 
However, compared to situations for which the relaxation dynamics is of a purely diffusive nature, 
application of the equilibrium identity (\ref{sumrule}) to an externally driven system represents 
a more severe approximation. 
For example, in the absence of an external potential field, equation (\ref{sumrule}) is automatically, and 
trivially, satisfied for a homogeneous and isotropic fluid. 
This is not the case for a driven system. 
Even when $V^{\rm ext}(\rb)=0$, the presence of an external flow field distorts the pair correlation functions and renders the integral term 
finite, whereas the spatial homogeneity of the one-body density yields a value of zero for the r.h.s.

There is considerable current research activity in the development of theoretical approaches to treat 
colloidal systems driven into nonequilibrium states by external flow. 
Much of the focus has been on the description of dense states close to the glass transition 
(see e.g. \cite{joeprl_08,pnas}). 
While the mode-coupling-type approaches employed in these studies are capable of capturing nonergodic 
behaviour, their application is restricted to systems with a spatially homogeneous density distribution. 
In contrast, Eq.(\ref{adv_ddft}) is ideally suited to address spatial inhomogeneity, but is incapable of 
describing glassy dynamics.

In the present work we will consider the application of (\ref{adv_ddft}) to driven steady states. 
In order to clearly expose the nature of the underlying approximations we will focus on the 
specific test case of interacting colloidal particles at a planar wall, with a shear flow acting parallel to the wall. 
Consideration of this particular external field and flow geometry reveals a serious deficiency of applying 
(\ref{sumrule}) to close the equation of motion for the one-body density of driven states. 
The physics which is lost in making the closure approximation arises from a coupling between the interparticle interactions and the external flow field and would, in an exact treatment, be captured 
implicitly by the flow induced distortion of $\rho^{(2)}(\rb_1,\rb_2)$.
Previous studies based on (\ref{adv_ddft}) have focused on two cases: (i) Noninteracting particles, for which the only relevant coupling is that between the external 
potential and flow fields \cite{rauscher1,mk_diploma}, 
(ii) Spherically inhomogeneous soft Gaussian particles \cite{dzubiella,rauscher2,rauscher3}. 
In these studies the combination of soft particle interactions and spherical inhomogeneity served to obscure the failings of Eq.(\ref{adv_ddft}) addressed in the present work.

The paper will be structured as follows:
In section \ref{system} we specify the system under consideration. 
In section \ref{coupling} we introduce the problem presented by the absence of a coupling between the external flow and the interparticle interactions. 
In sections \ref{mf_theory} and \ref{flowkernel} we develop a mean field theory which captures the desired coupling and propose a simple approximation for the required convolution kernel. 
In section \ref{excess} we detail the Rosenfeld functional used to approximate the excess free energy functional. 
In section \ref{results} we present and analyze the density profiles of hard-spheres at a hard wall, both in the presence and absence of gravity. 
In section \ref{discussion} we give a discussion and provide an outlook for future work.

\section{The System}\label{system}
We consider a system of $N$ spherical colloidal particles dispersed in an incompressible Newtonian solvent at temperature $T$. 
The diameter $d$ of the strongly repulsive colloidal core provides the basic 
unit of length. 
Assuming that the colloidal momenta are instantaneously thermalized, the time evolution of the probability distribution of particle positions, $\Psi(t)\equiv\Psi(\{ {\bf r}_i\},t)$, is dictated by the Smoluchowski equation \cite{dhont}
\begin{eqnarray}
\hspace*{1.1cm}\frac{\partial \Psi(t)}{\partial t} + 
\sum_{i} \bp_i\cdot {\bf j}_i =0,
\label{smol_hydro}
\end{eqnarray} 
where the probability flux of particle $i$ is given by
\begin{eqnarray}
{\bf j}_i={\bf v}_i(t)\Psi(t)
- \sum_{j} \D_{ij}\cdot(\bp_j - \beta\,{\bf F}_j)\Psi(t),
\label{flux}
\end{eqnarray}
with $\beta=1/k_BT$. 
The hydrodynamic velocity of particle $i$ due to the applied flow is denoted by 
${\bf v}_i(t)$ and 
the diffusion tensor $\D_{ij}$ describes (via the mobility tensor $\boldsymbol{\Gamma}_{ij}=\beta\D_{ij}$) 
the hydrodynamic mobility of particle $i$ resulting from a force 
on particle $j$.
The force ${\bf F}_j$ on particle $j$ is generated from the total potential energy 
according to ${\bf F}_j=-\bp_j U_N$ and includes the influence of an external one-body potential 
field, as well as the forces generated by interaction with other particles (taken here to be pairwise additive) 
\begin{eqnarray}
U_N(\{ {\bf r}_i\},t)
=\sum_{i}V^{\rm ext}(\rb_i,t) + \sum_{i<j}\phi(|\rb_i-\rb_j|).
\label{potential}
\end{eqnarray}
The three terms contributing to the flux thus represent the competing effects of (from left 
to right in (\ref{flux})) external flow, diffusion and potential interactions. 

In order to focus on the thermodynamic (as opposed to hydrodynamic) aspects of the cooperative particle 
motion we will neglect hydrodynamic interactions (HI) between the colloids. 
The expression (\ref{flux}) is thus approximated in two ways: (i) The influence of the $N$-particle 
configuration on the mobility of a given particle is neglected, $\D_{ij}=D_0\delta_{ij}$, 
where $D_0$ is the bare diffusivity. 
(ii) The velocity field is determined by the translationally invariant (traceless) velocity gradient 
tensor describing the affine motion, ${\bf v}_i={\bf v}({\bf r}_i,t)=\kap(t)\cdot\rb_i$. 
Neglecting the dependence of $\kap(t)$ upon the colloid configuration enables the externally applied flow 
to be prescribed from the outset, without requiring that this be determined as part of a self consistent 
calculation. 

\begin{figure} 
\includegraphics[width=8.5cm,angle=0]{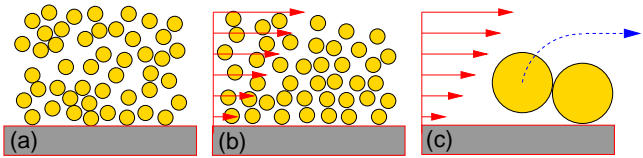}
\caption{
A schematic two-dimensional illustration showing the effect of shear flow on the microstructure 
of hard spheres at a hard wall. 
(a) A typical equilibrium configuration.  
(b) Shear flow leads to the formation of layers in the $xz$-plane as particles at different 
values of $y$ attempt to overtake each other. 
(c) Focusing on a binary collision close to the wall. The particle closer to the wall gets pushed against it, while the colliding particle is forced to `roll around' the other in order 
to move past.  
}
\label{figure1}
\end{figure}

\section{Interaction-flow coupling}\label{coupling}
In the present work we wish to study the influence of flow on a fully interacting, 
inhomogeneous system. 
So far, the only application of (\ref{adv_ddft}) has been to spherically inhomogeneous systems subject to a variety of flow fields \cite{rauscher1,mk_diploma,dzubiella,rauscher2,rauscher3} (representing a single colloid in a sea of ideal or Gaussian polymers). 
In particular, under shear flow, the ideal particles accumulate at the colloid surface within the two 
compressive quadrants ($I\equiv\rb\cdot(\kap+\kap^{T})\cdot\rb<0$) and are depleted from the extensional 
quadrants ($I>0$), leading to `wake' formation at larger shear rates \cite{mk_diploma}. 
The advected DDFT (\ref{adv_ddft}) thus captures a certain coupling between inhomogeneities in the density field and external flow. 
However, this `external potential-flow coupling' is a straightforward consequence of employing an external 
potential (e.g. a fixed sphere) which either directly obstructs the particles as they attempt to follow 
the affine flow, or perturbs the solvent flow field such that the particles are swept around the obstacle (when the appropriate Stokes flow is employed).

A more demanding and illuminating test of the advected-DDFT approach is provided by considering external 
potentials which do not directly hinder the affine path of the advected particles, but which may nevertheless 
be expected to generate flow dependent density profiles. 
Emphasis may thus be placed upon the nontrivial `interaction-flow coupling'. 
For the purpose of  the present work we will thus focus on the special case of a time-independent external 
potential which varies in $y$-direction and restricts the particles to move in the half space 
\begin{eqnarray}
V^{\rm ext}(\rb)=
\begin{cases}
\;\infty \hspace*{2cm}y<d/2  
\\
\;V^{\rm ext}_w(y) \hspace*{1.21cm}y>d/2. 
\end{cases}
\label{general_pot}
\end{eqnarray}
The translational invariance of $V^{\rm ext}_w$ within the $xz$-plane has the consequence that the resulting equilibrium density distribution varies in $y$-direction only. 
In addition to the external potential field, we specialize the external flow field to a steady simple shear with flow in $x$-direction and a linear gradient in the $y$-direction
\begin{eqnarray}\label{shear}
\vct{v}=y\,\dot\gamma\,\hat{\vct{e}}_x, 
\end{eqnarray}
with shear rate $\dot\gamma$ (corresponding to a velocity gradient tensor 
$\kappa_{\alpha\beta}=\dot\gamma\delta_{\alpha x}\delta_{\beta y}$). 
The pair potential entering (\ref{potential}) represents a hard-sphere repulsion
\begin{eqnarray}
\phi(r)=
\begin{cases}
\;\infty \hspace*{1.5cm}r<d  
\\
\;\;0 \hspace*{1.6cm}r>d. 
\end{cases}
\label{general_pair}
\end{eqnarray}
The situation under consideration is shown schematically in the second panel of Fig.\ref{figure1}. 
Note that the zero-shear plane may be located at $y=0$ without loss of generality, as only relative 
particle velocities are physically relevant.

Application of (\ref{adv_ddft}) to treat the specified nonequilibrium situation immediately reveals 
the problem at hand. 
We have already noted that our chosen geometry leads to translational invariance of the equilibrium density distribution within the $xz$-plane, $\rho_{\rm eq}(\rb,t)=\rho_{\rm eq}(y,t)$. 
For the shear flow (\ref{shear}) the advective term in (\ref{adv_ddft}) is thus given by
\begin{eqnarray}\label{theproblem}
\nabla\cdot(\rho(\rb,t)\vct{v}(\rb,t))=\nabla\cdot(y\,\dot\gamma\,\rho(y,t)\,\hat{\vct{e}}_x)=0. 
\end{eqnarray}
Within the advected-DDFT approach the applied shear flow has no influence on the density profile at the wall. 
Equation (\ref{adv_ddft}) thus reduces to (\ref{ddft}), which, for the present time-independent external 
potential, yields the equilibrium density profile. 
This disappointing conclusion contradicts physical intuition and presents a fundamental failing of 
Eq.(\ref{adv_ddft}). 

In the right panel of Fig.\ref{figure1} we sketch what we consider to be the correct physical picture. 
As a shear flow is applied parallel to the wall the particles experience a (nonconservative) force 
proportional to their perpendicular separation from the wall. 
Particles at different values of $y$ thus seek to move past each other, perturbing the equilibrium 
microstructure and leading, at higher shear rates, to the formation of particle layers in the 
$xz$-plane. 
In particular, when a pair of particles collide close to the wall, the particle at smaller  $y$ will 
be forced against the wall, whereas the second particle will be forced to roll around its neighbour 
in order to follow as closely as possible the affine solvent flow. 
These physical mechanisms should be manifest in the nonequilibrium density profile. 

We note that Brownian dynamics simulations \cite{brady_rev} display two dimensional particle layering
at intermediate shear rates, followed by an additional symmetry breaking in the vorticity direction 
at high shear rates, characterized by the formation of particle 
chains in $x$-direction. 
It is important to realize that the density distribution described by DFT represents an average over 
all $z$-coordinates of the particle chains (which, for an infinite system, are not pinned to a particular 
location in $z$). 
Our assumption that $\rho(\rb,t)=\rho(y,t)$ is thus fully justified for the present density functional based study.

\section{Mean field theory}\label{mf_theory}
In a system without HI, $N$-particle configurations for which colloidal particles overlap are forbidden 
by the infinitely repulsive colloidal pair potential.  
While an exact statistical mechanical treatment (i.e. exact evaluation of the integral 
in (\ref{sumrule})) would lend such unphysical configurations zero statistical weight, care must be exercised 
in approximate treatments which do not fully satisfy this important geometrical constraint. 
In the present situation it would appear that application of the equilibrium sum rule (\ref{sumrule}) 
does not satisfy exactly the no-overlap `core condition' when applied to driven states. 
However, it is not at all clear how to improve the approximation 
(\ref{sumrule}) in a way that both incorporates the flow induced distortion of $\rho^{(2)}(\rb_1,\rb_2,t)$ 
and enables its weighted integral to be approximated using an equilibrium free energy functional. 
For this reason we take an alternative route and attempt to include the missing physics by modifying the advective term in (\ref{adv_ddft}). 
This approach is motivated by considering the dynamics of hydrodynamically interacting 
dispersions. 

In a system with HI, the hydrodynamic velocity entering (\ref{flux}) can be decomposed into affine and particle induced fluctuation terms 
${\bf v}_i(t)=\kap(t)\cdot{\bf r}_i + {\bf v}^{\rm fl}_i(t)$, where ${\bf v}^{\rm fl}_i(t)$ can be 
expressed in terms of the third rank hydrodynamic resistance tensor \cite{lubrication}. 
The fluctuation term describes the disturbance of the affine solvent flow by the particles and ensures 
that a pair of approaching particles `flow around' each other, without coming into direct contact. 
The impenetrable character of  the particles, represented by the pair potential (\ref{general_pair}), thus 
enters indirectly by providing a boundary condition for the solvent flow. 
Integration of (\ref{smol_hydro}) over $N\!-\!1$ degrees of freedom yields an advective term   
\begin{eqnarray}\label{noproblem}
\nabla\cdot\, \rho(\rb,t) [\,\kap(t)\cdot\rb \,+\, \vb^{\rm fl}(\rb,t)\,] 
\end{eqnarray}
where $\vb^{\rm fl}(\rb,t)\equiv\langle \vb^{\rm fl}_i(t) \rangle$ is the conditional average over $N\!-\!1$ coordinates, given that a particle is located at $\rb$. 
In contrast to (\ref{theproblem}), the divergence in (\ref{noproblem}) is not neccessarily zero for the inhomgeneous system presently under consideration. 
For hydrodynamically interacting systems the fluctuation term may thus provide the desired contribution to the flux in $y$-direction.

The above considerations suggest a possible solution to  the problem posed by (\ref{theproblem}). 
By empirically incorporating a conditionally averaged fluctuation term into the velocity field driving our Brownian system,  
we aim to mimic the hydrodynamic fluctuation term discussed above.  
In this way we can correct, at least  to some extent, the occurance of unphysical overlaps by enforcing 
that there be no radial flux between pairs of particles at contact. 
We envisage a system of hard-core particles under shear flow in which there occur frequent and random binary 
collisions. At each collision the particles rotate around each other according to some specified rule 
(for our specific choice, see section \ref{flowkernel} below) before moving apart along their respective 
streamlines. 
For a homogeneous system, this mechanism gives rise to zero net flux through any given plane at constant $y$.  
However, in the presence of the external potential field (\ref{general_pot}), the inhomogeneous density distribution will lead to a $y$-dependent flux which will modify the equilibrium distribution, until it is 
balanced by an equal and opposite diffusive flux, thus establishing a nonequilibrium steady state. 

As the density profile under consideration varies only in $y$-direction, we need only seek the $y$-component 
of the fluctuation contribution. 
The arguments presented above thus suggest the mean field term  
\begin{eqnarray}\label{mf_term}
v^{\rm fl}_y(y,t)=\int_{-\infty}^{\,\infty}dy' \rho(y',t)\,\bar v^k_y(y-y')\label{eq:mf}
\end{eqnarray}
which expresses the contribution of the density at $y'$ to the average velocity at $y$. 
The `flow kernel' $\bar v^k_y(y)$ entering (\ref{mf_term}) describes the $y$-component of the velocity of a particle 
which collides with a neighbour at vertical separation $y$. 
Our modified version of (\ref{adv_ddft}) thus becomes
\begin{eqnarray}
\frac{\partial \rho(y,t)}{\partial t} &=& 
\frac{\partial}{\partial y}\Bigg[\,-\rho(y,t)v^{\rm fl}_y(y,t) \notag\\
&&\hspace*{0.5cm} +\;\;\; Pe^{-1}\rho(y,t)
\frac{\partial}{\partial y}\frac{\delta \mathcal{F}[\rho(y,t)]}{\delta \rho(y,t)} \Bigg],
\label{adv_ddft_mod}
\end{eqnarray}
where we have scaled length using $d$ and time using $\dot\gamma$, leading to an explicit appearance of the 
Peclet number, 
$Pe\equiv\dot\gamma d^2/D_0$, expressing the competition between affine advection and diffusive motion. 
The steady state solution of (\ref{adv_ddft_mod}) is given by 
\begin{eqnarray}
\rho(y)=z\exp\left(-\beta V^{\rm ext}(y) + c^{(1)}(y) + 
Pe\!\int_y^{\infty}\!\!\!dy' v^{\rm fl}_y(y')\right)\notag\\
\label{ss_solution}
\end{eqnarray}
where $z$ is the equilibrium fugacity and we have introduced the one-body direct correlation function 
\cite{bob_advances} 
\begin{eqnarray}
c^{(1)}(\rb)=-\frac{\delta \beta\mathcal{F}[\rho(\rb,t)]}{\delta \rho(\rb,t)}.
\end{eqnarray}
The integral in (\ref{ss_solution}) can be regarded as a nonequilibrium contribution to the intrinsic 
chemical potential.  
It should be noted that the requirement that a homogeneous density generates no particle flux in 
$y$-direction implies that the spatial integral over $v_y(y)$ is zero. 
For homogeneous states $v_y^{\rm fl}(y,t)$ is thus 
zero and the bulk chemical potential is unchanged from that in equilibrium.

\section{Flow kernel}\label{flowkernel}
In order to derive the flow kernel in Eq.~\eqref{eq:mf}, we consider the relative velocities of two particles, a tagged particle with position $\vct{r}_t$ and a reference particle at $\vct{r}_r$, during a scattering event, with $\Delta \vct{r}=\vct{r}_t-\vct{r}_r$. As discussed above, we neglect hydrodynamic interactions to keep the description as simple as possible. The incorporation of hydrodynamic interactions should in principle be possible in our approach. The following derivation of the scattering velocities is based on the assumption that in any situation, the particles adjust their velocities such that they minimize the friction with the solvent. The particles are at contact while they move around each other and then pass each other (see the right panel of Fig.\ref{figure1}). 
During the contact period, they have a nonvanishing velocity relative to the solvent. We will now derive the velocity $\vct{v}^k$ of the tagged particle in the frame comoving with the reference particle, which is assumed to move with constant velocity $\dot\gamma y_r$, i.e., we keep $y_r$ fixed. In a real scattering event one particle would move up and the other down. In our mean field picture, where the tagged particle moves in the fixed density distribution of other particles, we have to keep the $y$-position of the reference particle fixed. 
The velocity with minimal friction follows from minimization of the velocity relative to the solvent velocity $\vct{v}$, which we denote $\Delta {\bf v}$,
\begin{equation}
\Delta v^2=(\vct{v}^k-\vct{v})^2=(v^k_x-\dot\gamma \Delta y)^2+v_y^{k2}+v_z^{k2}.\label{eq:1}
\end{equation}
As long as the particles are in contact, they move on a trajectory with constant distance from each other, 
leading to 
\begin{equation}
\Delta r^2=d^2.
\end{equation}
Differentiation of this expression and elimination of $\Delta x$ leads to
\begin{equation}
v^k_x=-\frac{v^k_y\Delta y+v^k_z\Delta z}{\sqrt{d^2-\Delta y^2-\Delta z^2}}.
\end{equation}
Inserting in Eq.~\eqref{eq:1} yields
\begin{equation}
\Delta v^2=\left(-\frac{v^k_y\Delta y+v^k_z\Delta z}{\sqrt{d^2-\Delta y^2-\Delta z^2}}-\dot\gamma \Delta y\right)^2+v_y^{k2}+v_z^{k2}.\label{eq:11}
\end{equation}
We can now minimize this expression with respect to $v^k_y$ and $v^k_z$. This yields,
\begin{equation}\label{vy}
v^k_y (\Delta y, \Delta z)=-\dot\gamma \left(\frac{\Delta y}{d}\right)^2\sqrt{d^2-\Delta y^2-\Delta z^2}.
\end{equation}
In order to average this value over all possible values of $\Delta z$, we substitute
\begin{equation}
\Delta z=\sqrt{d^2-\Delta y^2}\sin \phi.
\end{equation} 
into (\ref{vy}) and integrate over $\phi$ to obtain (the factor of $\frac{1}{2\pi}$ is inserted as normalization, and the range of the integral is restricted to the half circle since the velocity is zero for $|\phi|>\pi/2$)
\begin{align}\label{vy1}
v^k_y (\Delta y)&=-\frac{1}{2\pi}\dot\gamma \left(\frac{\Delta y}{d}\right)^2 \sqrt{(d^2-\Delta y^2)}\int_{-\pi/2}^{\pi/2}\!\!\!\!d\phi \cos \phi\notag\\&=-\frac{\dot\gamma}{\pi} \left(\frac{\Delta y}{d}\right)^2 \sqrt{d^2-\Delta y^2}.
\end{align}
This result does not account for the density of particles at contact, relative to that in bulk. 
In order to include information about the local microstructure about a reference particle, the flow distorted inhomogeneous pair 
distribution function $g(\rb,y)$ should, in principle, be included as a prefactor in (\ref{vy}) before angular integration. 
Given that $g(\rb,y)$ is not known, we make the zeroth order approximation $g(\rb,y)= g_{\rm eq}(r)$, to arrive at our 
final result 
\begin{eqnarray}\label{kernel}
\bar v^k_y(\Delta y)=-\frac{\dot\gamma}{\pi}\, g_{\rm eq}(d) \left(\frac{\Delta y}{d}\right)^2 \sqrt{d^2-\Delta y^2}.
\end{eqnarray}

For hard-spheres the Carnahan-Starling expression $g_{\rm eq}(d)=(1-\phi/2)/(1-\phi)^3$ provides a simple and accurate expression for the contact value \cite{hansen}. 
For other choices of interaction potential $g_{\rm eq}(d)$ may be obtained using 
either integral equation methods \cite{brader_ijtp} or, more consistently, an equilibrium test-particle 
calculation employing the same Helmholtz free energy as that used to generate the dynamics. 

\section{Excess free energy}\label{excess}
Given the equation of motion (\ref{adv_ddft_mod}) and flow kernel (\ref{kernel}), we need to specify 
a particular approximation to the excess free energy functional in order to arrive at a closed theory 
for the density profile. 
For hard-sphere fluids the Rosenfeld functional \cite{rosenfeld} yields accurate results for both the 
microstructure and thermodynamics. 
Within the Rosenfeld approximation the excess free energy of hard-spheres is given by
\begin{eqnarray}
\mathcal{F}^{\rm hs}_{\rm ex}[\rho] &=& 
-n_0\ln(1-n_3)  
\,+\,\frac{n_1n_2-\boldsymbol{n}_1\cdot\boldsymbol{n}_2}{1-n_3} \notag\\
&&+\,\frac{n_2^3 - 3 n_2(\boldsymbol{n}_2\cdot\boldsymbol{n}_2)}
{24\pi(1-n_3)^2}.
\label{rosenfeldfunctional}
\end{eqnarray}
where the weighted densities are given by convolutions of the density profile 
\begin{equation}
n_{\alpha}(\mathbf{r})=\int d\mathbf{r'}\rho(\mathbf{r'})
\,\omega^{(\alpha)}(\mathbf{r}-\mathbf{r'}).
\label{weighted_density}
\end{equation}
The weight functions are characteristic of the geometry of the particles 
\begin{eqnarray}
\omega^{(3)}(\mathbf{r})&=&\Theta(r-R),\notag\\
\omega^{(2)}(\mathbf{r})&=&\delta(r-R),\notag\\
\omega^{(1)}(\mathbf{r})&=&\frac{\delta(r-R)}{4\pi R},\notag\\
\omega^{(0)}(\mathbf{r})&=&\frac{\delta(r-R)}{4\pi R^{2}},\notag\\
\boldsymbol{\omega}^{(2)}(\mathbf{r})&=&\frac{\mathbf{r}}{r}\delta(r-R),\notag\\
\boldsymbol{\omega}^{(1)}(\mathbf{r})&=&\frac{\mathbf{r}}{r}\frac{\delta(r-R)}
{4\pi R},
\label{weightfunctions}
\end{eqnarray}
where $R=d/2$ is the sphere radius. 
Although improved versions of the Rosenfeld functional do exist \cite{roth_review}, the original 
version \cite{rosenfeld} will prove sufficient for the present application. 


\section{Results}\label{results}

\subsection{Hard-spheres}
We first address the problem of hard-spheres at a hard wall ($V^{\rm ext}_w(y)=0$).  
The steady state equation (\ref{ss_solution}) was solved numerically (Picard iteration) using the flow kernel (\ref{kernel}) and the Rosenfeld approximation to the excess free energy. The contact value of the radial distribution function 
employed in (\ref{kernel}) was taken from the Carnahan-Starling  equation of state \cite{hansen}. 

\subsubsection{Intermediate shear rates}
\begin{figure}[!t] 
\includegraphics[width=8.2cm,angle=0]{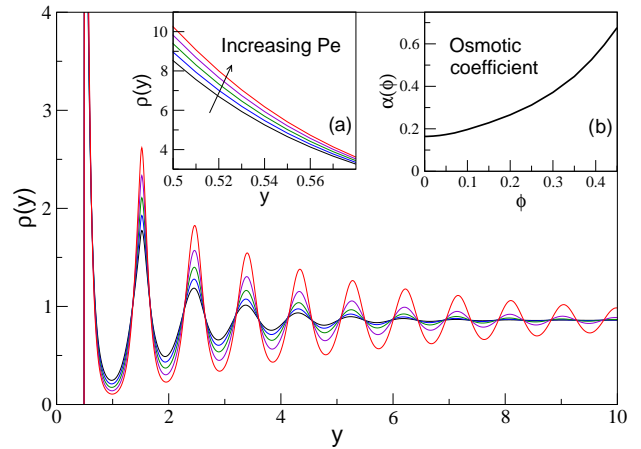}
\caption{
Steady state density profiles of a hard-sphere fluid at volume fraction $\phi=0.45$ for 
$Pe=0, 3.142, 6.283, 9.425$ and $12.566$. 
As the Peclet number is increased the oscillatory structure of the profile becomes more 
pronounced, reflecting the formation of particle layers in the $xz$-plane. 
Inset (a) focuses on the region close to the wall, where the contact value $\rho(d/2)$ 
increases linearly with $Pe$. 
Inset (b) shows the density dependence of the coefficient determining the nonequilibrium 
contribution to the reduced osmotic pressure 
$\beta\Pi_{\rm ne}(\phi,Pe)=\alpha(\phi)\phi^2 Pe$, as determined from the contact value.  
}
\label{figure2}
\end{figure}

In Fig.\ref{figure2} we show density profiles calculated for a volume fraction $\phi=0.45$ at 
various (low to intermediate) values of the Peclet number. 
In equilibrium, $Pe=0$, the density profile shows a typical oscillatory structure arising from local 
packing contraints at the wall. 
Applying a finite $Pe$ leads to an increase in both the contact value (see inset (a)) and 
height of the oscillatory peaks, which is accompanied by an increase in the depth of the minima. 
The enhanced structure of the profile is a direct consequence of the collision mechanism built into the advective term of our theory (c.f. Fig.\ref{figure1}.c) and indicates the development of particle layers in nonequilibrium steady states at finite $Pe$. 
Despite the highly structured character of the nonequilibrium profiles, it should be noted that the 
adsorption (i.e. the spatial integral of $\rho(y)-\rho_b$) remains independent of $Pe$, where $\rho_b=\frac{6\phi}{\pi d^3}$ is the bulk colloid density. 
While this is a straightforward consequence of the continuity equation underlying (\ref{adv_ddft}), it 
nevertheless provides a useful check for our numerical results. 

\subsubsection{Osmotic pressure}
For hard-spheres at a hard-wall, the contact value of the density profile satisfies the sum rule
\begin{eqnarray}
\beta \Pi = \rho(d/2),
\label{osmotic}
\end{eqnarray}
where $\Pi$ is the osmotic pressure.  
While the sum rule (\ref{osmotic}) is generally applied to equilibrium, there seems to be no 
reason why it should not be equally valid for the present nonequilibrium situation 
(although, as far as we are aware, there currently exists no mathematical proof of this assertion). 
Our numerical calculations show that, for a given volume fraction, the contact value $\rho(d/2)$ 
increases linearly over a the entire range of Peclet numbers investigated ($Pe=0\rightarrow 10$ for each 
volume fraction considered). Employing the sum rule (\ref{osmotic}) we thus find that 
the numerically obtained osmotic pressure obeys the following relation 
\begin{eqnarray}
\beta \Pi(\phi,Pe) = \beta \Pi_{\rm eq}(\phi) + \alpha(\phi)\phi^2 Pe
\label{osmotic_fit}
\end{eqnarray} 
where $\alpha(\phi)$ is a volume fraction dependent coefficient. 
\begin{figure}[!t] 
\includegraphics[width=8.0cm,angle=0]{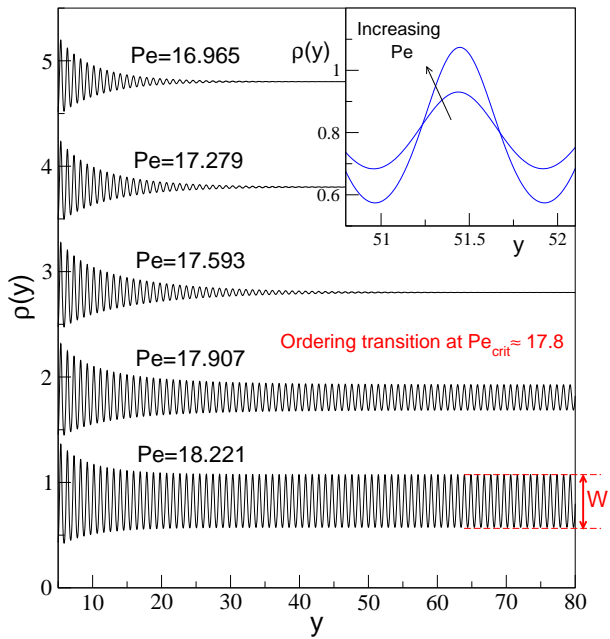}
\caption{
Steady state density profiles for volume fraction $\phi=0.42$ at $Pe=16.965, 17.279, 
17.593, 17.907$ and $18.221$. 
For clarity the profiles have been translated vertically. 
As $Pe$ is increased the profiles display an increasingly slow oscillatory decay to the bulk density. 
At a critical value of $Pe=Pe_{\rm crit}$ the Brownian motion is no longer able to restore the equilibrium structure far from the wall and shear effects dominate. 
For $Pe>Pe_{\rm crit}$ the oscillations no longer decay and the entire sample enters a layered state, 
characterized by a well defined oscillation amplitude. 
The width of the laning oscillations away from the wall (and which extend throughout the entire sample) defines an order parameter $W$ characterizing the nonequilibrium transition. 
The inset focuses on a single peak ($\phi=0.42$) within the layering region for $Pe=17.907$ and 
$18.221$. The layering peaks can be well approximated by a Gaussian.   
}
\label{figure3}
\end{figure}

The definition of $\alpha(\phi)$ in the second term of (\ref{osmotic_fit}) is motivated by the exact low volume 
fraction results of Brady and Morris \cite{brady_morris}. By solving exactly the pair Smoluchowski equation 
in the low density limit for hard-spheres without HI it has been shown that the osmotic pressure 
(obtained from the trace of the stress tensor) is given by 
\begin{eqnarray}
\beta \Pi(\phi\rightarrow 0,Pe) = \rho_b + \frac{4}{3\pi^2}\phi^2 Pe
\label{brady_osmotic}
\end{eqnarray} 
In inset (b) of Fig.\ref{figure2} we show the volume fraction dependence of $\alpha$. 
Gratifyingly, the fact that $\alpha$ exhibits a low volume fraction plateau confirms that the present theory 
indeed captures the correct scaling ($\sim \phi^2 Pe$) of the flow induced correction to the 
osmotic pressure. 
The fact that we recover the correct low density scaling is a nontrivial output of our approach. 
However, the limiting value $\alpha(\phi\rightarrow 0)=0.164$ obtained in the present work differs 
from the exact value of $4/3\pi^2=0.135$ by a factor of $1.2$. 
Given the rather severe approximations employed in the present work, namely the mean-field term (\ref{mf_term}) and flow kernel (\ref{kernel}), it should not be surprising that there is some deviation from the exact result. 
Nevertheless, the recovery of the correct low density scaling is reassuring and suggests that performing calculations 
with a renormalized Peclet number $Pe^*=Pe/1.2$ may be appropriate, should a detailed comparison with simulation or experiment be required. 

We note that time-dependent solutions of (\ref{adv_ddft_mod}) (not considered in the present work) would enable study 
of the transient behaviour of the Osmotic pressure resulting from time-dependent changes in the applied shear flow, 
e.g. the onset of steady shear flow \cite{zausch}.

\subsubsection{Laning transition}  
Turning now to higher values of $Pe$, we show in Fig.\ref{figure3} density profiles for $\phi=0.42$ and $Pe=16.695 \rightarrow 18.221$, focusing on the layering structure away from the direct vicinity of the wall (the region $y>5$ is shown). 
As $Pe$ is increased from zero to values around $17.6$ the oscillatory structure shows an increasingly slow decay with distance from the wall, indicating that Brownian motion is gradually succumbing to the influence of the applied shear flow. 
For $Pe>17.6$ the decay length of the oscillatory profiles shows a strong sensitivity to variations in the 
Peclet number and diverges at a critical 
value $Pe_{\rm crit}\approx 17.8$.  
This divergence signifies the onset of an ordered phase for which the asymptotic density profile is 
characterized by a well defined period and amplitude of oscillation. 

The emergence of an infinitely extended oscillatory profile at a critical value of the Peclet number is a 
nontrivial prediction of the present theory and signifies a non equilibrium transition to an inhomogeneous steady state. 
Such layered states have been observed in Brownian dynamics simulations \cite{foss_brady} but have thus far remained 
inaccessable to microscopically based theories. 
For $Pe>Pe_{\rm crit}$ it is of interest to look at the structure of the individual oscillations within the layered phase.
In the inset to Fig.\ref{figure3} we show a single density peak at $y\approx 51.4$ for $Pe=17.907$ and $18.221$.  
For larger values of $Pe$ the peak becomes both narrower and higher, reflecting the reduced influence 
of Brownian motion, which acts to damp the oscillations and restore equilibrium. 
The density peaks in the layering region may be well approximated by a Gaussian, implying 
the existence of two-dimensional particle planes at high $Pe$ values, with harmonic restoring forces acting against 
random out-of-plane fluctuations.

Shear induced layering phases, similar to those predicted by the present theory, have been observed in both colloidal experiments \cite{hoffman1,hoffman2,ackerson_clark,ackerson_pusey,ackerson} and 
Brownian dynamics simulations of hard-sphere dispersions 
\cite{foss_brady}. More recently, experiments on noncolloidal dispersions (no Brownian motion) under oscillatory shear 
have shown that the presence of irreversible processes when the particles collide can give 
rise to self-organization and the formation of particle lanes \cite{pine}. 

\subsubsection{Phase diagram}
\begin{figure}[!t] 
\includegraphics[width=8.0cm,angle=0]{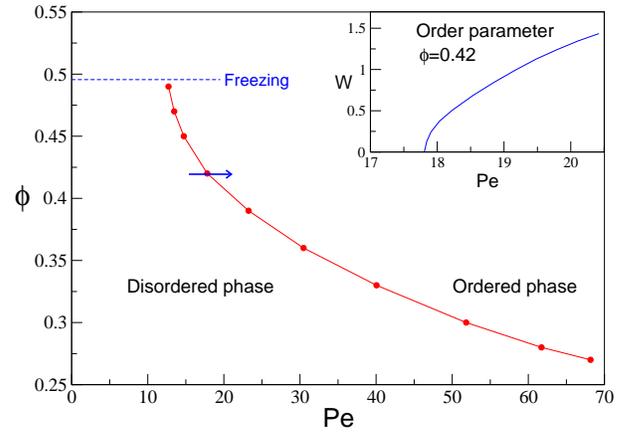}
\caption{
The phase boundary in the $(\phi\,,Pe)$ plane separating the disordered phase from the ordered `laning' phase (lines serve as a guide for the eye between calculated data points). 
The hard-sphere freezing transition at $\phi=0.494$ is indicated by the broken line. 
The inset shows the order parameter $W$ as a function of $Pe$ for $\phi=0.42$, following the path 
indicated by the blue arrow in the main panel. 
Above $Pe_{\rm crit}$ the numerical data suggests a continuous transition with the order parameter 
varying as $W\sim(Pe\!-\!Pe_{\rm crit})^{\frac{1}{2}}$ for small $Pe\!-\!Pe_{\rm crit}$.  
}
\label{figure4}
\end{figure}

The oscillation amplitude of the density in the limit $y\rightarrow\infty$ serves as an order parameter 
characterizing the nonequilibrium transition from a locally layered state, homogeneous in bulk, to a 
fully macroscopic layered phase. 
Specifically, $W\equiv\max(\rho(y\rightarrow\infty))-\min(\rho(y\rightarrow\infty))$ provides a suitable order parameter (see Fig.\ref{figure3}). 
In Fig.4 we show the nonequilibrium phase diagram in the $(\phi\,,Pe)$ plane, obtained by examination of $W$ as 
a function of $Pe$.  
For each volume fraction density profiles were calculated on a large grid extending beyond $300$ particle diameters. 
For $Pe<Pe_{\rm crit}$ the converged profiles clearly decay to zero as a function of $y$, well 
within our sample size (as for the profiles for $Pe=16.965, 17.279$ and $17.593$ in Fig.\ref{figure3}). 
For $Pe>Pe_{\rm crit}$ iteration of Eq.(\ref{ss_solution}) results in a `laning region' which grows out from the wall indefinitely until the laning structure covers the entire range of the numerical grid.  
The value of $W$ for laning states may thus be operationally defined as the density difference between the 
minina and maxima of the oscillations at a distance sufficiently far removed from the wall. 
In practice, $W$ was estimated from the variation of the profile around $y=150$. 
The inset to Fig.\ref{figure4} shows $W$ as a function of $Pe$ for $\phi=0.42$, following the path indicated by an arrow in the main figure. For values of $Pe$ close to, but above, the transition, $W$ is well described by a square 
root.

The phase diagram shown in Fig.\ref{figure4} is consistent with that calculated in Brownian dynamics simulations of charge stabilized colloidal dispersions (see Fig.4 in \cite{grest}), provided that the temperature used in the simulation study is identified as an inverse volume fraction. 
In \cite{grest} temperatures both above and below the equilibrium freezing transition were considered and the nonequilibrium order-disorder phase boundary found to vary continuously through the freezing transition. 
In the present study we prefer to restrict 
ourselves to volume fractions below freezing ($\phi_{\rm fr}=0.494$) in order to avoid the possible complications 
which may arise from the presence of underlying metastable states. 
A serious study of the complex interaction between crystal nucleation and external flow goes beyond the scope 
of the present work. 

Finally, we note that the value $Pe_{\rm crit}=14.7$ obtained from the present theory for $\phi=0.45$ is remarkably 
consistent with Brownian dynamics simulations performed at the same volume fraction 
(c.f. Figure.3 in \cite{foss_brady}). 
The simulations predict that a layered structure emerges within the range $Pe=10 \rightarrow 30$.

\subsubsection{Bulk laning}
The results presented in the previous section indicate that the presence of the dynamical mean field 
term in (\ref{ss_solution}) gives rise to an instability with respect to laning when the Peclet number exceeds a certain critical value. 
Although we have concentrated on the particular problem of particles at a hard-wall, it would appear that the density 
inhomogeneities induced by the wall simply serve to `seed' the generation of a laning structure for 
$Pe>Pe_{\rm crit}$. 
It may thus be anticipated that for supercritical values of $Pe$, any kind of density fluctuation, regardless of its 
amplitude, will be sufficient to initiate laning. 

In order to test the above hypothesis we have solved (\ref{ss_solution}) for a range of $Pe$ numbers using the following initial guess for numerical iterative solution 
\begin{eqnarray}\label{initial}
\rho_{\rm init}(y)=\rho_b + a\exp(-b(y-y_0)^2),
\end{eqnarray}
for various values of the parameters $a, b$ and $y_0$. 
For $Pe<Pe_{\rm crit}$ the perturbation is eroded during the iteration procedure and yields the steady state solution $\rho(y)=\rho_b$ for all values of the parameter set $(a,b,y_0)$.
For $Pe>Pe_{\rm crit}$ any finite value of the parameter $a$ is sufficient to seed the laning and a fully laned steady state solution, extending over the entire numerical sample length, is obtained, regardless of the values of $b$ and $y_0$ employed. 
In this sense, it would appear that, for supercritical states, any amount of `numerical dirt' in the initial homogeneous density distribution is sufficient to generate a fully laned steady state. 
Moreover, we have confirmed that the values of $Pe_{\rm crit}$ thus obtained are entirely consistent with the phase boundary shown in Fig.\ref{figure4}, which was calculated in the presence of a hard-wall.
Given the above observations it would be of interest to perform a fully time-dependent solution of (\ref{adv_ddft_mod}). 
Such a calculation, which we defer to a future publication, would also enable predictions to be made regarding 
the timescale upon which lanes are formed and its dependence upon the supersaturation $Pe\!-\!Pe_{\rm crit}$.

\begin{figure}[!t] 
\includegraphics[width=8.0cm,angle=0]{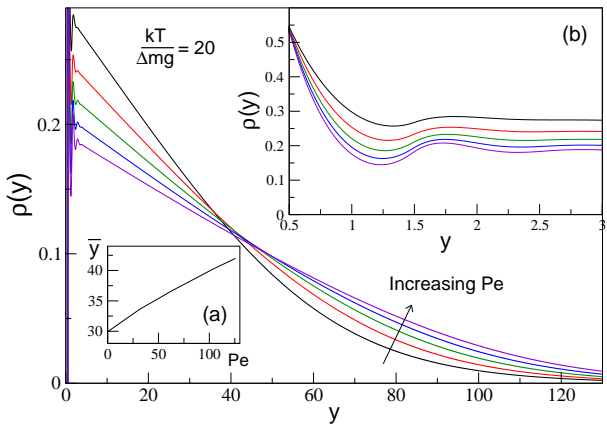}
\caption{
Steady state sedimentation profiles of a relatively dilute dispersion for $Pe=0,31.416,62.832,103.673$ and 
$125.664$ at a fixed value of the gravitational length $k_BT/g\Delta m=20$. 
The black curve corresponds to $Pe=0$ and is obtained from a static DFT calculation 
at a fugacity $z=1.5$. 
Due to the conservation equation underlying our DDFT, steady state profiles at finite 
$Pe$ have the same adsorption as the profile for $Pe=0$, i.e. particle number is conserved. 
Inset (a) shows the center-of-mass $\bar{y}$ (see Eq.\ref{com} as a function of $Pe$. 
Inset (b) focuses on the region close to the wall and demonstrates fact that the contact value is 
independent of $Pe$. 
}
\label{figure5}
\end{figure}

\subsection{Influence of gravity}
We now consider adding an extra component to the external potential 
\begin{eqnarray}
V_w^{\rm ext}(y)=yg\Delta m, 
\end{eqnarray}
where $\Delta m$ is the bouyant mass of a colloid and $g$ is the gravitational 
field strength. We thus address the influence of the shear flow (\ref{shear}) upon colloidal 
sedimentation profiles. 
Choosing a fugacity $z=1.5$ and gravitational length $\xi=k_BT/g\Delta m=20$ yields the equilibrium sedimentation profile shown in Fig.\ref{figure5}, 
for which the local volume fraction remains rather low, even in the vicinity of the wall. 
As $y$ increases, the local packing oscillations give way to a monotonic decrease of the density. 
It may thus be anticipated that for finite values of $Pe$, the flow kernel built into our mean-field 
theory will lead to a net transport of particles from regions of high density to regions of lower 
density until the scattering flux is balanced by the gravity-biased diffusion of particles towards 
smaller $y$ values.  

The expectation of flow induced broadening of the sedimentation profiles is confirmed by the steady 
state results shown in Fig.\ref{figure5}. 
Note that the particle number (i.e. area under each of the curves in Fig.\ref{figure5}) is conserved 
and is independent of $Pe$. 
The canonical nature of the DDFT imposes that the broadening of the sedimentation profile with increasing 
$Pe$ is accompanied by an overall decrease in the local density within the range $0<y<40$. 
It is interesting to consider this change of the density distribution in view of the recently discussed violation of the fluctuation dissipation theorem (FDT) \cite{Krueger09,Berthier02} in sheared systems. This violation was expressed in terms of the fluctuation dissipation ratio $X_{\rm FDT, f}$ defined as the ratio of response and thermal fluctuations for observable $f$ (see, e.g. \cite{Krueger09} for details). In equilibrium, one has $X_{\rm FDT,f}=1$, while under shear, $X_{\rm FDT,f}<1$ is observed. Since, by definition, the ratio $X_{\rm FDT, f}$ is proportional to $(k_BT)^{-1}$ (when keeping response and fluctuations $T$-independent), one can also describe the FDT violation in terms of an effective temperature $T_{\rm FDT,f}=T/X_{\rm FDT, f}$ which in turn is larger than $T$ \cite{Berthier02}. (Note that the dependence of $X_{\rm FDT, f}$, and hence $T_{\rm FDT,f}$, on observable $f$ is unclear.)
Here, we are tempted to define in analogy the center-of-mass ratio $X_{\rm com}$,
\begin{eqnarray}\label{comr}
X_{\rm com}=\frac{\bar{y}(0)}{\bar{y}(\dot\gamma)},
\end{eqnarray}
with $\bar{y}(\dot\gamma)$  the center-of-mass of the density distribution at shear rate $\dot\gamma$,
\begin{eqnarray}\label{com}
\bar{y}=\frac{\int_0^\infty \!dy\, y \rho(y)}{\int_0^\infty \!dy\, \rho(y)}.
\end{eqnarray}
At low density, $\rho(y)\propto \exp[-y/\xi]$ and $\bar{y}=\xi\propto k_BT$, independent of shear. At higher density, $\rho$ in Fig.~\ref{figure5} does not follow a simple exponential, but one still expects $\bar{y}(0)\propto k_BT$, as long as packing effects are not too dominant. This confirms the close analogy of our defintion of $X_{\rm com}$  to $X_{\rm FDT,f}$: if for the system under shear, the ratio $X_{\rm com}$ will be smaller than unity, one can formally interpret this in terms of an effective temperature larger than $T$.
Inset (a) to Fig.5 shows that the center-of-mass under shear is indeed larger then in equilibrium, i.e., we indeed have $X_{\rm com}<1$ in accordance to the findings for the ratio $X_{\rm FDT,f}$.  The decrease of $X_{\rm com}$ as function of shear rate resembles the behavior of $X_{\rm FDT,f}$, which was also found to decrease with shear rate \cite{Krueger09,Berthier02}. We furthermore expect that $X_{\rm com}$ decreases with density and note $X_{\rm com}\to 1$ for $\phi\to0$ (where $\bar{y}(\dot\gamma)\to \xi$), as observed for $X_{\rm FDT,f}$. The center-of-mass ratio hence shows the same overall properties as the fluctuation-dissipation-ratio. This suggests that both are driven by similar physical processes. These findings are also interesting in view of efforts towards a thermodynamic definition of an effective temperature of the system under shear \cite{Langer07}. We realize that a comparison of the concrete values of $X_{\rm com}$ and $X_{\rm FDT,f}$ is not possible since the system under gravity is different from the bulk-systems studied in \cite{Krueger09,Berthier02,Langer07}, as the density depends both on $y$ and $\dot\gamma$.  Future work on a single tagged heavy particle in a bath of density matched particles might prove more useful in this respect.

Despite the broadening of the profiles as a function of $Pe$, the ultimate asymptotic decay can always be fitted by a Boltzmann decay, $\rho(y\rightarrow\infty)\sim \exp(-y/\xi)$. This is expected since  the density far away from the wall is low and $\varrho(y)$ hence approaches the Boltzmann-distribution.

The inset to Fig.\ref{figure5} focuses on the region $0.5<y<3$. 
Despite the major changes in density distribution induced by the applied shear flow, it is striking that 
the contact value $\rho(d/2)$ remains independent of $Pe$, in contrast to our previous findings at $g=0$ (see Figs.\ref{figure2} and \ref{figure3}). 
This is not surprising. 
For any finite $g$, the steady state density profile has an adsorption $\Gamma\equiv\int_0^{\infty}\!dy\,\rho(y)$ 
corresponding to the average number of particles in a column in $y$-direction with unit area in the $xz$-plane. 
In a gravitational field this column of particles thus exerts a force $\Gamma g \Delta m$ on the wall and determines the contact value of the density distribution. 
As particle number is conserved within the DDFT, it follows that the contact value of the density at the wall 
will be independent of $Pe$, as observed in our numerical solutions. 
The fact that the variation in contact value as a function of $Pe$ is distictly different for the two cases 
$Pe=0$ and $Pe\ne 0$ is related to the singular behaviour of $\rho(y\rightarrow\infty)$ in the limit $g\rightarrow 0$.

\begin{figure}[!t] 
\includegraphics[width=8.2cm,angle=0]{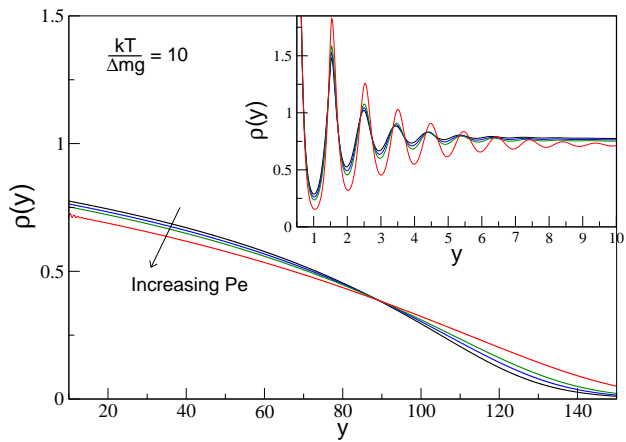}
\caption{
Steady state sedimentation profiles of a dense dispersion for $Pe=2,4,6$ and $15$ and at separations removed from the wall. The inset shows the local structure close to the wall. Although some local layering can be induced close to the wall at high shear rates, 
the gravitational force suppresses the development of an extended layering phase. 
As in Fig.\ref{figure5}, the contact values remains independent of $Pe$.      
}
\label{figure6}
\end{figure}

Fig.\ref{figure6} shows further sedimentation profiles for a state with $\xi=10$ and larger local volume fraction than those shown in Fig.\ref{figure5}. 
As previously, the profiles broaden with increasing $Pe$. 
Due to high local density in the vicinity of the wall, it may be expected that the layering transition identified in our calculations at $g=0$ may become relevant at sufficiently high $Pe$ values. 
The profile at the highest $Pe$ value shown in Fig.\ref{figure6} ($Pe=15$) does indeed show the development of a layering structure close to the wall, similar to that in Fig.\ref{figure2}. 
However, the gravitational force acting on the particles supresses the development of long range oscillations and disrupts the formation of a macroscopic layering phase at any finite value of $Pe$. 
As for the profiles shown in Fig.\ref{figure5}, the contact value at the wall remains independent of $Pe$ over the range of parameters investigated.

\section{Discussion}\label{discussion}
We have applied dynamical density functional theory to calculate the density 
profiles of a colloidal liquid at a wall under shear flow. The chosen flow geometry served to highlight 
failings of the existing DDFT approach to driven states and a semi-empirical correction was proposed 
to reintroduce the missing physical mechanism. Calculations performed at various volume fractions and 
Peclet numbers have revealed that the new approximation captures a non-equilibrium phase transition to 
an ordered laning state, for shear rates above a critical value of $Pe$. Moreover, sedimentation profiles 
are dramatically altered by the application of shear flow, which leads to an increase in height of 
the colloidal center-of-mass with increasing shear rate. 
The behavior of the center-of-mass ratio $X_{\rm com}$ is in 
qualitative agreement with the previously studied 
fluctuation-dissipation-ratio under shear. 
The study of a single tagged heavy       
particle in a bath of density matched particles might be an interesting variant for the future.

The mean-field correction to the advection term is presently rather empirical in character, arrived at using 
physical arguments, and it would be desirable 
to place this on a more rigorous basis, ideally as part of a systematic scheme for improving the DDFT under 
external flow. Whether this is possible remains to be seen. 
In some sense, the present state of the theory for driven states is reminiscent of the early days of 
equilibrium DFT, for 
which the first attempts to go beyond the local density or square gradient approximation relied on the introduction 
of empirical weight functions to incorporate spatial nonlocality \cite{bob_review}. 
The insight gained from these studies proved very useful for the development of subsequent nonlocal approaches with a better 
foundation in statistical mechanics. We thus hope that the present work may provide stimulus for further developments in applying DDFT to driven nonequilibrium states. 

The physical 'scattering' mechanism built into the present theory generates a nontrivial coupling between density inhomogeneities and external flow, but has no effect on systems with a homogeneous density distribution. 
While this is likely to be appropriate for certain colloidal systems, it may represent an approximation for others. 
Imposing shear flow on a homogeneous dispersion generally leads to the development of finite normal stresses, 
which have been associated with the phenomenon of shear-induced particle migration \cite{morris_rev}. 
It is thus conceivable that a shear-induced drift of particles to regions of lower shear rate could result in a density gradient through the sample. 
Very recent experiments on PMMA hard-sphere-like colloids suggest that flow-concentration coupling can lead to 
a novel form of shear-banding \cite{mike_banding}. 
However, the banding reported in \cite{mike_banding} only occurs for volume fractions above the glass transition, 
whereas the present work is focused purely on colloidal liquid states. 
 
Both the standard form of DDFT (\ref{ddft}) and its advected extension (\ref{adv_ddft}) are based on an implicit adiabatic assumption which neglects the time taken for the (one-time) pair correlation functions to 
equilibrate following a change in the average density profile. It may thus be anticipated that in very dense 
systems, for which the structural $\alpha$-relaxation time becomes important, the pair correlation functions will 
be unable to keep up with changes in the density, leading to a breakdown of the adiabatic approximation. 
The fact that the structural relaxation time of driven dense states is determined by the inverse flow rate 
$\dot\gamma^{-1}$ (at least for states with $\dot\gamma^{-1}<\tau_{\alpha}$) raises the interesting possibility 
that the adiabatic approximation may be more accurate when applied to calculate 
the response of dense systems to time-dependent changes in flow rate than to time-dependent changes in external potential. 
The present work has focused on steady state response and the next step in our research program will be to extend 
our studies to treat time-dependent shear flow. 

An important simplification of the present treatment is that hydrodynamic interactions have been neglected. 
This excludes from the outset the development of the hydrodynamically bound clusters which may form at 
very high shear rates and which have been suggested as a possible mechanism for shear thickening \cite{joe_review}. 
As we focus here on the low and intermediate Peclet numbers characteristic of the onset of shear thinning, this 
omission should not be too severe. 
More fundamental is the fact that the ordered phase observed in Brownian dynamics simulations \cite{foss_brady} and captured by the present theory is apparently absent in Stokesian dynamics simulations including the full solvent hydrodynamics \cite{foss_brady2}. 
It would thus appear that hydrodynamic interactions can render the ordered phase unstable. 
Nevertheless, we consider it important that any prospective theory of driven colloids be able to descibe first 
the simpler case of interacting Brownian particles, before seeking to refine this to include hydrodynamics at some level. While it may well be that the (approximate) incorporation of hydrodynamic interactions into the theory disrupts the laning behaviour reported here, we can at least be sure that such an improved theory has a sound physical basis and that the laning observed in Brownian dynamics \cite{foss_brady} will indeed emerge should we choose to switch-off the hydrodynamics.  
Such a gradual theoretical development, adding new physical aspects step-by-step, is important in developing a 
robust theory and tackling the fully hydrodynamic problem from the outset would be unlikely to deliver this.\\

\section*{Acknowledgements}
It is a pleasure to contribute the present work to a Festschrift celebrating the achievements 
of Professor Evans. One of us (JMB) had the privilege to complete a Ph.D. under Bob's 
supervision. His enthusiastic and creative approach to physics have made a deep and lasting 
impression.
The authors would like to thank Matthias Fuchs for numerous discussions on the subject 
of colloid dynamics, we both benefitted from the stimulating environment in the Konstanz 
Soft Matter Theory Group. 
Funding was provided by the SFB-TR6, the Swiss National Science Foundation and the 
Deutsche Forschungsgemeinschaft under grant number KR 3844/1-1.

{}


\begin{thebibliography}{}

\bibitem{bob_advances}
R.~Evans, Adv.Phys. {\bf 28} 143 (1979); 

\bibitem{bob_review}
R.~Evans, in {\em Fundamentals of inhomogeneous fluids}, edited by 
D.~Henderson (Dekker, New York, 1992).

\bibitem{rosenfeld}
Y.~Rosenfeld, 
Phys.Rev.Lett. {\bf 63} 980 (1989).

\bibitem{colpol}
M.~Schmidt, H.~L\"owen, J.M.~Brader and R.~Evans, 
Phys.Rev.Lett. {\bf 85} 1934 (2000).

\bibitem{colpol1}
R. Evans, J. M. Brader, R. Roth, M. Dijkstra, M. Schmidt, and H. L\"owen,
Philos. Trans. R. Soc. London, Ser. A {\bf 359}, 961 (2001). 

\bibitem{onsager}
J.M Brader, A. Esztermann and M. Schmidt, 
Phys.Rev.E {\bf 66} 031401 (2002). 

\bibitem{marconi1}
U.~Marini Bettolo Marconi and P.~Tarazona,  
J.Chem.Phys. {\bf 110} 8032 (1999). 

\bibitem{marconi2}
U.~Marini Bettolo Marconi and P.~Tarazona,  
J.Phys.: Condens.Matter {\bf 12} A413 (2000).

\bibitem{archer}
 A.J.~Archer and R.~Evans, J.Chem.Phys. {\bf 121} 4254 (2004). 

\bibitem{finken}
Garnet Kin-Lic Chan and R.~Finken,  
Phys.Rev.Lett. {\bf 94} 183001 (2005). 

\bibitem{espanol}
P.~Espanol and H.L\"owen, J.Chem.Phys. {\bf 131} 244101 (2009). 

\bibitem{marconi3}
U.~Marini Bettolo Marconi and S.~Melchionna, 
J.Chem.Phys. {126} 184109 (2007). 

\bibitem{rex1}
M.~Rex and H.~L\"owen, 
Phys.Rev.Lett. {\bf 101} 148302 (2008).

\bibitem{rex2}
M.~Rex and H.~L\"owen,
Eur.Phys.J.E {\bf 28} 139 (2009). 

\bibitem{rauscher_hydro}
M. Rauscher, 
J.Phys.: Condens.Matter {\bf 22} 364109 (2010).

\bibitem{rex3}
M.~Rex, H.H.~Wensink and H.~L\"owen, 
Phys.Rev.E {\bf 76} 021403 (2007). 

\bibitem{wensink} 
H.H.~Wensink and H.~L\"owen, 
Phys.Rev.E {\bf 78} 031409 (2008). 

\bibitem{rauscher2}
M.~Rauscher, A.~Dominguez, M.~Kr\"uger and F.~Penna, 
J.Chem.Phys. {\bf 127} 244906 (2007).

\bibitem{joeprl_08}
J.M. Brader, M.E. Cates and M. Fuchs,
Phys.Rev.Lett. {\bf 101} 138301 (2008).

\bibitem{pnas} 
J.M.~Brader, Th.~Voigtmann, M.~Fuchs, R.G.~Larson and M.E.~Cates,
Proc. Natl. Acad. Sci. U.S.A. {\bf 106} 15186 (2009).

\bibitem{rauscher1}
M.~Kr\"uger and M.~Rauscher, 
J.Chem.Phys. {\bf 127} 034905 (2007).

\bibitem{mk_diploma}
M.~Kr\"uger, Diploma Thesis, University of Stuttgart, Germany 2006. 

\bibitem{dzubiella}
F.~Penna, J.~Dzubiella and P.~Tarazona, 
Phys.Rev.E {\bf 68} 061407 (2003)..

\bibitem{rauscher3}
C. Gutsche, F. Kremer, M.~Kr\"uger, M.~Rauscher, R. Weeber and J. Harting,  
J.Chem.Phys. {\bf 129} 084902 (2007).


\bibitem{dhont}
J. K. G.~Dhont, {\em An introduction to the dynamics of colloids}
(Amsterdam, Elsevier, 1996).

\bibitem{lubrication}
S.~Kim and S.J.~Karilla, {\em Microhydrodynamics, principles and selected applications} 
(Butterworth-Heinemann, Boston, 1991).

\bibitem{hansen}
J.-P.~Hansen and I.R.~McDonald. {Theory of Simple Liquids} 
(Academic, London, 1986).

\bibitem{brader_ijtp}
J.M.~Brader, Int.J.Thermophys. {\bf 27} 394 (2006). 

\bibitem{roth_review}
R. Roth, J.Phys.:Condens.Matter {\bf 22} 063102 (2010). 

\bibitem{brady_morris}
J.F.~Brady and J.F.~Morris, J. Fluid Mech. {\bf 348} 103 (1997).

\bibitem{hoffman1}
R.L. Hoffman, Trans.Soc.Rheol. {\bf 16} 155 (1972). 

\bibitem{hoffman2}
R.L. Hoffman, J.Colloid Interface Sci. {\bf 46} 491 (1974).

\bibitem{ackerson_clark}
B.J. Ackerson and N.A. Clark, Phys.Rev.Lett. {\bf 46} 123 (1982).   

\bibitem{ackerson_pusey}
B.J. Ackerson and P.N. Pusey, Phys.Rev.Lett. {\bf 61} 1033 (1988).
 
\bibitem{ackerson}
B.J. Ackerson, J.Rheol. {\bf 34} 553 (1990). 

\bibitem{foss_brady}
D.R. Foss and J.F. Brady, J.Rheol. {\bf 44} 629 (2000).

\bibitem{pine}
L. Cort\'e, P.M. Chaikin, J.P. Gollub and D.J. Pine, 
Nature Physics, {\bf 4} 420 (2008). 

\bibitem{zausch}
J.~Zausch, J.~Horbach, M.~Laurati, S.U.~Egelhaaf, J.M.~Brader, T.~Voigtmann and M.~Fuchs,
J.Phys.:Condens.Matter {\bf 20} 404210 (2008).

\bibitem{brady_rev}
J.F.~Brady, Chem.Eng.Sci. {\bf 56} 2921 (2001). 

\bibitem{joe_review}
J.M. Brader, J.Phys.:Condens.Matter {\bf 22} 363101 (2010). 

\bibitem{grest}
W. Xue and G.S. Grest, Phys.Rev.Lett. {\bf 64} 419 (1990). 

\bibitem{morris_rev}
J.F.~Morris, Rheol.Acta {\bf 48} 827 (2009). 

\bibitem{mike_banding}
R. Besseling, L. Isa, P. Ballesta, G. Petekidis, M.E. Cates and W.C.K. Poon, 
arXiv:1009.1579 (2010). 

\bibitem{foss_brady2}
D.R. Foss and J.F. Brady, J.Fluid.Mech. {\bf 407} 167 (2000). 

\bibitem{Krueger09}
 M. Kr{\"u}ger and M. Fuchs, Phys. Rev. Lett. {\bf 102}, 135701 (2009).
\bibitem{Langer07}
J. S. Langer and M. L. Manning, Phys. Rev. E {\bf 76}, 056107 (2007).
\bibitem{Berthier02}
L. Berthier and J.-L. Barrat, J. Chem. Phys. {\bf 116}, 6228 (2002).

\end{thebibliography}
\end{document}